\documentclass[
aps,%
12pt,%
final,%
notitlepage,%
oneside,%
onecolumn,%
nobibnotes,%
nofootinbib,%
superscriptaddress,%
noshowpacs,%
centertags]%
{revtex4}

\DeclareMathAlphabet{\mathds}{U}{dsrom}{m}{n}
\newcommand*{\myeqref}[1]{(\ref{#1})}
\newcommand*{\okr}{ \mathds{S} }
\newcommand*{\rea}{\mathds{R}}
\newcommand*{\Bd}{\mathop{\mathrm{Bd}}\nolimits}
\newcommand*{\Cl}{\mathop{\mathrm{Cl}}\nolimits}
\newcommand*{\scri}{\mathcal{J}}
\newcommand*{\ssy}[5]{#1,   #2  \textbf{#3}, #5  (#4)\rlap{.}}

\begin{document}
\title{``TOPOLOGICAL CENSORSHIP'' IS NOT PROVEN}

\author{\firstname{S.~V.}~\surname{Krasnikov}}
\email{gennady.krasnikov@pobox.spbu.ru}
\affiliation{%
The Central Astronomical Observatory at Pulkovo, St.~Petersburg,
Russia}%
 %
 %
 %
 %


\begin{abstract}
I show that there is a  significant lacuna in the proof of the
theorem known as ``Topological Censorship'' (a theorem forbidding a
solution of Einstein's equations to have some topological features, such as
traversable wormholes, without violating the averaged null energy
condition). To fill the lacuna one would probably have to revise the class of
spacetimes for which the theorem is formulated.
\end{abstract}

\maketitle

\section{Introduction}

In their known paper \cite{fsw} Friedman, Schleich, and
 Witt (FSW) formulated a theorem called ``Topological Censorship'':

\noindent \textbf{Theorem \cite{fsw}.} If an asymptotically flat, globally
hyperbolic spacetime $(M, g_{ab})$ satisfies the averaged null energy
condition (ANEC), then every causal curve $\gamma $ from ${\scri}^-$ to
${\scri }^+$ is deformable to $\gamma _0\ {\rm rel}\ {\scri}$.

\medskip

\noindent Here the term ``asymptotically flat'' implies\footnote{The
``compactness'' required in \cite{fsw} from
    $\widetilde
    M$
    is \emph{actually} some ``spatial compactness'' and the derivatives
    $\tilde \nabla_a\tilde \nabla_b \Omega$ are \emph{actually}
    assumed to vanish on $\scri$ (FSW, unpublished). So, asymptotical flatness in
    \cite{fsw} is understood in the sense of \cite{gh}.}, in
particular,
 that there is a
conformal completion $(\widetilde M, \tilde g_{ab})$ where $\widetilde
M$ is a spacetime with $\tilde g_{ab} =\Omega^2 g_{ab}$ for some
$\Omega$ that vanishes on the boundary, ${\scri} = {\widetilde M} - M$,
which is a disjoint union of past and future parts, ${\scri}^+ \cup
{\scri}^-$, each
 having the topology $\okr^2\times\rea $ with the $\rea$'s complete null
generators. And $\gamma_0$ is a timelike curve with past endpoint in
${\scri}^-$ and future endpoint in ${\scri}^+$ that lies in a simply
connected neighborhood $U$ of ${\scri}$.

The theorem serves  as a basis for a few further
 theorems \cite{cw,galloway,jv}, but what seems even more important
 is its role in the studies of the traversable wormholes. Already in their
 pioneering work \cite{Tho} Morris and  Thorne cited an argument, due to
 Page, suggesting that the existence of
 such wormholes is closely linked with the Weak energy condition
 violations (which shaped the direction of research in this field for decades).
 ``Roughly speaking, the reason is that bundles of
light rays (null geodesics) that enter the wormhole at one mouth and
emerge from the other must have cross-sectional areas that initially
decrease and then increase. [This] requires negative energy density'' \cite{Tho}. The problem with this powerful, and as it proved, very useful
argument is that \emph{rigorously} speaking it is incorrect. Consider, for
example, the null rays emanated towards the center from every point of a
spacelike two-sphere in Minkowski space. The bundle formed by these
rays  initially has a decreasing cross-sectional area, but later (after the rays
pass through the center) it becomes increasing, even though the space is
\emph{empty}. So, in the general case the conversion from decreasing to
increasing does \emph{not} require WEC violations. To make the
statement more rigorous Morris and  Thorne reformulated it in different
terms: ``A roughly spherical surface on one side of the wormhole throat,
from the viewpoint of the other side, is an ``outer trapped surface'' ---
which, by Proposition~9.2.8 of Ref.~22\footnote{That is our
Ref.~\cite{HawEl}.}, is possible only if the ``weak energy condition'' is
violated'' . But \emph{this is not true either}. Proposition~9.2.8 of
\cite{HawEl} forbids a remote observer to see an outer trapped surface
only if the spacetime is regular predictable and, as a consequence,
``asymptotically empty and simple'' . This latter condition cannot be
relaxed:  the proof leans on the fact that a certain set
 ${\scri}^+ \cap {\dot J}^+({\cal P}, \bar {\cal M})$ --- let us denote it $S_{9.2.8}$ ---
 is non-empty, and this fact
 is \emph{proven} in \cite{HawEl}, see Lemma 6.9.3, only for asymptotically
 empty and simple spaces. At the same time we know that a
 spacetime with a wormhole \emph{cannot} be asymptotically
 empty and simple: there are null geodesics that come from infinity and
 enter the wormhole never to leave it, while in asymptotically
 empty and simple spaces by definition every null  geodesic in $M$ must have
 two endpoints on  $\partial M$.

 Thus, an important mathematical problem arose pertinent to the
 wormhole physics: what is a (sufficiently broad) class of spacetimes in
 which the presence
 of a wormhole would imply the WEC violation? The ``topological
 censorship'' theorem offered a solution and a very attractive
 one: the class of globally hyperbolic asymptotically flat spacetimes
 comprises most of physically interesting cases. However the proof of the
 theorem contains, as we shall see in the next section, a serious lacuna.
 Curiously enough, the fact that is left unproven is exactly the same as in Page's
reasoning: the set appearing in the displayed equation \myeqref{A} is essentially $S_{9.2.8}$
 mentioned above.

The proof of the  ``topological censorship'' theorem offered in \cite{fsw}
contains the fatal gap consisting  in the fallacy of the following implication:
``[\dots] if ${\scri}_{\alpha }^+ \cap {J}^+({\cal T})$ is both closed and
open, then ${\scri}_{\alpha }^+$ is disconnected''. The aim of the present
note is to state this fact, but also to clarify the role of the false implication
and to show that be it correct there would be good reasons to consider the
theorem valid. To that end in the next section I outline a possible proof of
the theorem. It may differ slightly from the original one \cite{fsw}, because
the latter is rather vague at some points (and a thorough analysis of those
points   would take one far beyond the scope of this paper). So, it should be
stressed that no possible demerit of the following reasoning can refute our
main point: the topological censorship theorem is not proven and will
remain unproven until its proof leans on the false implication cited above.

Whether the \emph{claim} of the theorem is true is yet to be found (it is
imaginable, for example, that   the conditions of the theorem exclude the
possibility of \myeqref{eq:zvezd}). Meanwhile, one can use weaker results.
For example, one can prove the theorem under the \emph{additional
assumption} that the inclusion \myeqref{eq:zvezd} does not
hold\footnote{This, in fact, is already done, see Theorem~6.1 in
    \cite{KK}.}.  Sometimes such results are also referred to as ``topological
censorship'' . They, however, are very different --- both mathematically and
physically --- from the theorem considered in this paper and their
discussion is beyond its scope.
\section{The lacuna}

   Assume there
       \emph{is} a   $\gamma$
       non-homotopic to $\gamma_0$.  Let  $\cal M$ be the  universal
       covering of $M$ and
       $\widetilde{\cal M}$ be a conformal completion of  $\cal M$. The
       boundary of $\widetilde{\cal M}$
       consists of disjoint sets ${\scri}_{\alpha }^\pm$ and
       there is a causal curve  $\Gamma_0$ (a lift of   $\gamma_0$) with the
       end points in some  ${\scri}_{0}^-$ and   ${\scri}_{0}^+$. The existence of
  $\gamma$ then implies that there is also a
causal curve $\Gamma'$ (a lift of   $\gamma$) from  ${\scri}_{0}^-$ to
${\scri}_{\alpha_1}^+$, where $\alpha_1\neq 0$ (instead of $\Gamma'$
FSW choose to consider another lift of   $\gamma$ --- that with the future
end point in ${\scri}_0^+$; it is denoted by $\Gamma$).
  On its way from $\scri_{0}^-$,     $\Gamma' $  meets a ``very large sphere''       ${\cal T}\in U $, which is defined to be  a smooth closed
      orientable two-surface with
      the following property:  one of the two null future directed
      congruences orthogonal to  ${\cal T}$ --- denote this congruence
      $\mathfrak C_1$ --- terminates at    ${\scri}_{0}^+$
      and the second one --- denoted $\mathfrak C_2$  --- has   negative
      expansion in all points of   ${\cal T}$.

To show that, in fact, such a $\Gamma' $ (\emph{i.~e.,}  a causal curve
connecting     ${\cal T}$ to ${\scri}_{\alpha_1}^+$) cannot exist, and thus
to prove the whole theorem it would suffice to find a point
\begin{equation}\label{A}
p\in
{\scri}_{\alpha_1 }^+\cap\Cl _{\widetilde{\cal
M}}(\mathcal H ), \qquad \mathcal H \equiv \Bd_\mathcal{M}J^+({\cal T})
\end{equation}
(throughout the paper I write $\Bd_YX$ and $\Cl_YX$ for, respectively,
the boundary and the closure of $X$ in $Y$). Indeed, $\cal M$, being a
covering of the globally hyperbolic $M$, is globally hyperbolic. Hence, for
any point $p_i\in \mathcal H $ there is a null geodesic segment (a
generator of  $\mathcal H $) which ends in $p_i$, starts in ${\cal T}$, and
is disjoint with $I^+ ({\cal T})$. By continuity a segment with the same
properties | denote it $\lambda $ | would have to exist for $p$ too.
Evidently $\lambda $ must be orthogonal to ${\cal T}$, but as FSW argue,
see Lemma 1, it does not belong to $\mathfrak C_1$. Nor can it belong to
$\mathfrak C_2$ [the negative initial expansion in combination with
ANEC (recall also that $\lambda $ is future-complete in  $\cal M$) would
ensure in such a case \cite{borde} that there is a point $q\in \lambda $
conjugate to ${\cal T}$. Beyond $q$, $\lambda$ cannot remain in
$\mathcal H $,  it enters $I^+({\cal T})$]. A contradiction.

The existence of $p$ can be established by examining an appropriately
chosen subset $A$ of $\scri_{\alpha_1}^+$. In \cite{fsw} this set  is taken
to be
\[
A\equiv\scri_{\alpha_1}^+ \cap J^+({\cal T}),
\]
but $J^+({\cal T})$ (understood as a subset of $\cal M$ (not of
$\widetilde{\cal M}$), which is necessary, in particular, to guarantee its
closedness) is disjoint with $\scri_{\alpha_1}^+$, so I think it more
rigorous to define
\[
A\equiv\scri_{\alpha_1}^+ \cap \Cl_{\widetilde{\cal M}}J^+({\cal T}).
\]
The thus defined $A$ is non-empty (due to $\Gamma' $) and its
non-openness in $\scri_{\alpha_1}^+$  would, indeed, imply the existence
of $p$, because any neighborhood of a point $r\in \Bd_
{\scri_{\alpha_1}^+ } A$ must contain points of $J^+({\cal T})$, but also
some points of ${\cal M} - J^+({\cal T})$. Hence $r$ is in
$\Cl_{\widetilde{\cal M}}(\mathcal H)$ and can be taken as $p$.

Thus, to prove the theorem it remains to prove that $A$ is not open in
$\scri_{\alpha_1}^+$. And FSW argue (see Lemma 2 in \cite{fsw}) that
$A$ cannot be open, because it is closed, so be it at the same time open,
$\scri_{\alpha_1}^+$ would have to be disconnected (while it is connected
by definition). This argument, however, is \emph{false}. A set \emph{can}
be closed, open, and connected all at the same time. Such will be $A$ if $
A=\scri_{\alpha_1}^+$, \emph{i.~e.,} if
 \begin{equation}
\scri_{\alpha_1}^+ \subset \Cl_{\widetilde{\cal M}}J^+({\cal
T}).\label{eq:zvezd}
\end{equation}
And in the general case no reasons are seen to exclude such a possibility.

\end{document}